\documentstyle[seceq]{ptptex}
\title{%        %You can use \\ for explicit line-break
Trans-Planckian redshifts 
and the substance of the space-time river
}
\author{%       %Use \sc for the family name
Ted Jacobson\footnote{E-mail address: jacobson@physics.umd.edu}
}

\inst{%         %Affiliation, neglected when [addenda] or [errata]
Department of Physics, Univerity of Maryland\\
College Park, MD 20742-4111 USA}

\recdate{%      %Editorial Office will fill in this.
}

\abst{%       %this abstract is neglected when [addenda] or [errata]
Trans-Planckian redshifts in cosmology and outside
black holes may provide windows on a hypothetical
short distance cutoff on the fundamental degrees of freedom.
In cosmology, such a cutoff seems to require 
a growing Hilbert space, 
but for black holes, Unruh's sonic analogy has given rise to 
both field theoretic and lattice models demonstrating
how such a cutoff in a fixed Hilbert space 
might be compatible with a low energy
effective quantum field theory of the Hawking effect.
In the lattice case, the outgoing modes arise via
a Bloch oscillation from ingoing modes.
A short distance cutoff on degrees of freedom
is incompatible with local Lorentz
invariance, but may nevertheless be compatible with general
covariance if the preferred frame is defined non-locally
by the cosmological background. Pursuing these ideas in a 
different direction,
condensed matter analogs may eventually allow for 
laboratory observations of the Hawking effect. 
This paper introduces
and gives a fairly complete but brief 
review of the work that has been done in these
areas, and tries to point the way to some 
future directions.
}

\begin{document}

\maketitle

\def\k{\kappa}
\def\d{\delta}

\section{Introduction}
 
It was inspiring to be in Kyoto with the purpose of looking
forward toward the next century of gravitational physics.
The present century gave us general relativity, with its
profound and beautiful understanding of gravitation. In this
view, we can think of spacetime as a river, and of gravitation
as the inhomogeneity of the flow. At the YKIS99 meeting we 
heard about vortices, waves, bifurcations, and stones in the
river. The subject of my talk was a question about the substance
of the river itself: the ancient question whether space and time
are continuous or discrete. 

The ultraviolet divergences of quantum field theory and the
infinite curvature singularities of general relativity
call for a fundamental short distance cutoff of some kind.
Perhaps spacetime is locally discrete, or perhaps locality
itself is not valid, as string theory suggests. But to  
learn something about the physics of the cutoff we must find
a way to access that remote territory.  

The modern viewpoint\cite{Weinberg:1996kw} is that quantum
field theory is an effective description of collective 
degrees of freedom  of an underlying medium whose nature remains
unknown, and need not be known, in order to do physics below
some cutoff momentum scale. 
However, 
there are two familiar settings in which the usual separation
between long and short distance scales breaks down, and the 
short distance physics is, potentially, unmasked.
These are the expansion of the universe, and the event horizon 
of a black hole. In both cases there is a tremendous, trans-Planckian,
redshift.

The redshift at a black hole 
horizon means that the low energy effective
field theory is not self-contained, since 
low energy outgoing modes evolve from degrees of freedom 
above the cutoff. 
The same is true in cosmology, since the expansion
of the universe redshifts degrees of freedom from above to below the 
effective field theory cutoff. In cosmology, the puzzle seems
to be even deeper than for a black hole, since 
the number of 
degrees of freedom must grow as the universe
expands if there is a fundamental short distance cutoff. 
In the black hole problem, by contrast, there is a time 
translation symmetry of the black hole spacetime, and 
translational symmetry is broken, which may
allow for a solution to the puzzle not involving 
a time dependence of the number of underlying degrees 
of freedom. 

It would seem natural to address the trans-Planckian redshift
puzzle in string theory, since in this theory
there are no degrees of freedom localizable below the 
string scale. Unfortunately,
the remarkable success in understanding 
the process of Hawking radiation of closed string states
by near-extremal D-branes\cite{Das:1997ta}
seems to be of no help in this endeavor,  since it
concerns processes at weak coupling in a flat background
spacetime.  The challenge is to account for the evolution
of the ingoing degrees of freedom into the
outgoing modes (and therefore the Hawking radiation) in the 
presence of a horizon.
To address this challenge would 
require an understanding of localized observables 
in a black hole background
at strong coupling, an understanding not presently available
in string theory. 

The situation may turn out to be better in the 
framework of the ``holographic" AdS/CFT duality\cite{magoo}, 
which is an outgrowth
of D-brane theory. If true, this duality would 
solve both the UV divergence and curvature singularity
problems, the former because the CFT is finite, and the latter
because the CFT lives on a non-dynamical background spacetime
and has no gravity in it! Although the description
of localized bulk observables in terms of observables of
the dual boundary CFT remains
an elusive quest, with no sign of rapid progress in sight,
perhaps the existence of an in-out mapping can be understood
without solving this problem. I leave such a possibility
for future work, and concentrate in this article on
other approaches to understanding the trans-Planckian puzzle.

Rather than reproducing the content of my talk at YKIS99, 
most of which is anyway covered more completely in published
articles\cite{CJlattice,JMlattice}, 
I have taken the opportunity of this contribution
to step back and present a broader view of the whole subject.
The underlying ideas and motivations are discussed here, and 
I have made an effort to review all of the related work of which
I am aware. I have also tried to look forward and discuss 
possibilities for future work that might be worthwhile and important.
 
The remainder of this paper is organized as follows. 
The cosmological and black hole trans-Planckian
redshift puzzles are discussed in section 2.
In section 3 the sonic analogy is introduced,
and the dispersive field theory models to which it gave rise
are discussed. Section 4 is devoted to lattice models,
and section 5 discusses the various condensed matter
analogs that have been proposed so far.
Finally, section 6 addresses the issue of Lorentz 
non-invariance. 

\section{Trans-Planckian redshift puzzles}

\subsection{Cosmological redshift}
In a pioneering paper written in 1985 (in a field that
does not yet exist), Weiss\cite{Weiss} pointed out serious
obstacles for the problem of hamiltonian lattice
quantum field theory in an expanding universe, which I 
will now summarize.
If the lattice points are co-moving with the cosmological
expansion, their proper separation grows in time, which 
leads to difficulties to be discussed momentarily.
One can instead place the lattice points on trajectories
that maintain their proper separation as follows.

In a spatially flat Robertson-Walker spacetime, for example,
with line element $ds^2= dt^2 - a(t)^2(dr^2 + r^2d\Omega^2)$,
the new coordinate $\rho=a(t)r$ casts the 
line element in the form
$ds^2= (1-H^2\rho^2)dt^2+2H\rho d\rho dt-d\rho^2 -\rho^2d\Omega^2$,
where $H=d(\ln a)/dt$ is generally a function of time.
If lattice points were to sit at fixed
$(\rho,\theta,\phi)$, their proper separation at fixed 
$t$ would be constant. However, since the vector 
$\partial/\partial t$ is spacelike for $\rho>H^{-1}$,
the Hamiltonian generating this ``time-translation" would be 
unbounded below. Weiss suggested this problem could be avoided with 
a boundary condition at $\rho=H^{-1}$, but noted that this
does not look like a promising approach, since the boundary
condition is in general time-dependent. The nature of the surface
$\rho=H^{-1}$ depends on the function $a(t)$.\cite{rho} 
For $a(t)=t^n$, the surface is spacelike if $n<1/2$,
outgoing lightlike if $n=1/2$ and timelike if $n>1/2$.
In any case, information flows into the region 
$\rho<H^{-1}$ from the boundary, and one doesn't know what
this information should be without solving the dynamics 
outside the boundary. In the spacelike case one could 
impose the data as an ``initial" condition, however since the
boundary is not a surface of constant $t$ one would not know
what condition is appropriate without solving the dynamics outside
the boundary. 

In the case of deSitter space where
$H$ a constant, the surface $\rho=H^{-1}$ is 
ingoing lightlike, and is in fact just the deSitter horizon.
Information cannot cross this horizon from outside to inside,
so it plays the same role for the inside as a black hole
horizon plays for the exterior of a black hole. But even in 
this case there is a difficulty, since the acceleration
of the lattice points at fixed $(\rho,\theta,\phi)$
diverges as the horizon is approached. 
According to the discussion in section \ref{lattice} 
on lattice black holes, I would expect this infinite
acceleration to lead to unphysical behavior of the lattice theory.

If the lattice points are instead co-moving with the expansion,
their proper separation grows in time. To maintain
small enough lattice spacing for a calculation spanning several
e-foldings of the scale factor, one must begin with
an exponentially small spacing. Moreover, since the density
of states is then time dependent, the bare parameters
of the theory must be functions of the cosmic time in order 
to keep the renormalized parameters fixed at a fixed proper scale. 
It would seem terribly impractical to keep track
of all the degrees of freedom and continuously adjust the
bare couplings so that the renormalized couplings are constant.

Much more natural would be to somehow add lattice points as 
the universe expands, to maintain a constant proper density
of points. This is not straightforward however,
since it is not 
unitary evolution on a fixed Hilbert space. Nevertheless,
if there is a fundamental short distance cutoff,
some such generalization of quantum theory seems unavoidable. 
It also seems desirable, because it might 
neatly account for the low initial entropy of the universe, 
since the actual number of degrees of freedom would have been very 
small at the beginning.

The idea of a growing cosmological 
Hilbert space has been discussed in
a few places I am aware 
of\cite{osgoodhill,unruhtime,pablomora}, and 
a proposal for how it might emerge from quantum gravity
is discussed in \cite{pablomora}.
The idea of a changing Hilbert space in general is 
natural in the consistent histories formulation of quantum
mechanics, as has been emphasized by Hartle\cite{Hartle}. 
The unitary inequivalence of the 
Hilbert space at different times 
for a quantum field in curved spacetime 
has recently been formulated by 
Anastopoulos\cite{Anastopoulos}
in terms of a time-dependent Hilbert space
using the histories formalism. It also seems that the algebraic
formulation of quantum field theory\cite{haag} 
has room for what would
amount to a time dependent Hilbert space.

A growing Hilbert space was introduced
for practical reasons in a continuum-based calculation of 
Ramsey and Hu\cite{Ramsey}. 
They numerically
evolved the scale factor coupled to a self-interacting 
quantum scalar field 
in inflationary semiclassical cosmology, using 
a $1/N$ approximation. For economy of computation,  
at any given time in the evolution
only modes up to some fixed large proper cutoff wavevector 
were included, the higher modes being physically irrelevant.
This meant that, as the universe 
expanded, entirely new modes entered the Hilbert space of 
the calculation. These modes were inserted in their 
instantaneous adiabatic vacuum state. This sort of scheme
was previously suggested as an effective description
of cosmology with a growing Hilbert space\cite{ultrashort}.

\subsection{Black hole horizon redshift}

If the short wavelength outgoing modes near the event horizon
of a black hole are in their ground state (as defined by
the time of an infalling observer), then these modes will be 
occupied at the Hawking temperature at infinity. By ``short
wavelength" and ``near" here I mean compared with the 
size of the black hole, as measured by an infalling observer.
Whatever the details of the theory at ultrashort distances,
as long as this outgoing vacuum condition is met, the
Hawking effect will occur.\cite{TJcutoff} 

The question that interests me, however, is by what mechanism
can a theory with a cutoff produce these outgoing modes at all?
If they can be accounted for,  the fact 
that they are delivered in their ground state would
be an important but presumably straightforward consequence
of the fact that the vacuum at short distances---whatever 
it is---remains unexcited as it propagates in the
black hole background, which is characterized by much longer
distance and time scales.       

If you are more practically than philosophically inclined,
consider the problem the following way. Suppose you want to numerically
calculate what comes out of an evaporating black hole, in
an interacting quantum field theory, using a lattice 
formulation. How can the lattice theory with, say,
Planck scale lattice spacing, possibly 
produce the outgoing modes, when it clearly cannot ``store"
them in a trans-Planckian reservoir at the horizon? 

When I first started thinking about this problem I suspected
that, as in the cosmological case, it would be necessary to 
generate new states in the Hilbert space, corresponding to the
outgoing modes. However, as stated 
earlier, the black hole is stationary and breaks translational
symmetry, so the analogy with
cosmology is not perfect. Indeed, it seems that there 
is a way to produce the outgoing modes within the confines
of a fixed Hilbert space. 
        
\section{Sonic analogy and dispersive field theory models}
A condensed matter analogy, Unruh's sonic black hole 
analogy\cite{Unruh81,Unruh95},
led the way to a possible resolution of the puzzle
of the origin of the outgoing black hole modes.\footnote{A
very different suggestion for how to avoid the need for fundamental
high frequency degrees of freedom, using ``superoscillations",
is discussed in Refs. \cite{Rosu} and \cite{ReznikSO}.}
I say ``a possible" resolution rather than 
``the" resolution, since the proposed scenario requires
a kind of modification of fundamental short distance physics
which has not been incorporated into any complete theory,
let alone been confirmed experimentally.

\subsection{Sonic analogy}
\label{sonic}
 
The idea of Unruh's analogy is that when  fluid flows faster than 
the speed of sound somewhere in an inhomogeneous flow, 
a sonic horizon appears which is in many ways analogous to a
black hole horizon. The sound field---perturbations of the
fluid flow---behaves in fact precisely like a massless 
relativistic field in a curved background spacetime
determined by the flow parameters,
provided the flow is irrotational, 
barotropic and inviscid.\cite{Unruh81,Moncrief,Visser}
For our present purposes the essential point is independent
of these details, and is just the fact that the sonic 
horizon is an infinite redshift surface for sound waves.
This poses a ``trans-Bohrian" puzzle, since a real physical
fluid cannot support sound of arbitrarily short wavelength.
In particular, at wavelengths smaller than the intermolecular
spacing the effective field theory of sound becomes invalid
and the density of states drops to zero. Nevertheless,
in the presence of such a sonic horizon there must indeed
be outgoing sound modes, for these are just some of the 
collective degrees of freedom of the fluid that must exist.

So how does a real fluid manage to produce the outgoing modes
at a sonic horizon? Since the sound speed is the
top speed for excitations of the fluid, modes cannot propagate
from inside to outside the horizon. 
The only possibility that conserves the number of degrees
of freedom of the fluid is therefore that the outgoing modes
come from ingoing degrees of freedom that are somehow 
turned back at the horizon. (There can be no talk of creating
degrees of freedom in a fluid, which is surely in principle
a self-contained quantum system.) Though this initially
sounds rather strange, especially given that a horizon
is normally thought to swallow everything that approaches it,
it turns out to be an example of a general phenomenon
that occurs for dispersive waves in an inhomogeneous medium.

Clearly modes cannot be ``reflected" from the horizon, since 
there is no sharp interface at which reflection could take place.
Rather, the reversal of group velocity must happen continuously,
as a result of smooth evolution from one branch of the dispersion
relation to another. The dispersion relation relating
frequency $\omega$ to wavevector $k$ for an atomic fluid
has the form $\omega=c_s k$ for wavelengths long compared to
the interatomic spacing, but as the wavevector grows this 
is modified and the group velocity drops lower than 
the long wavelength speed of sound $c_s$. In superfluid
Helium-4, for example, the group velocity actually drops to zero
and then reverses sign before the so-called roton minimum is reached. 

Due to this drop of group velocity,
an outgoing mode traced backwards in time in the WKB approximation
will asymptotically approach a point {\it outside} the sonic horizon, 
where the group velocity and flow velocity are equal 
and opposite.\cite{ultrashort}
What happens at this point is that the 
group velocity continues to drop, and the wavepacket reverses direction 
and propagates back away from the horizon.\cite{Unruh95} 
In other words, now viewing
the process forward in time, an ``outgoing" short wavelength mode
with low group velocity is dragged in toward the horizon by the 
faster fluid flow. As this happens the wavevector decreases and the
group velocity increases, eventually reaching and then exceeding the
flow velocity, at which point the wavepacket begins propagating
back out away from the horizon. 

\subsection{Dispersive field theory models}
It is not necessary to solve the many-body problem of a real atomic
fluid in order to explore this phenomenon.
Unruh\cite{Unruh95} studied the scenario just described by 
numerically solving the wave equation
for a 1+1 dimensional free field theory,
with higher spatial derivative terms designed 
to produce a dispersion
relation of the form 
\begin{equation}
\omega=k_0 \bigl(\tanh(k/k_0)^n\bigr)^{1/n}
\label{tanh}
\end{equation}
in the co-moving frame. The wavevector $k_0$ sets the scale
at which the deviations from the massless wave equation
become important. 
Unruh confirmed the reversal of group velocity, and found that
in the process there is some conversion from the 
positive to the negative frequency branch of the dispersion relation,
in just the right amount to yield the Hawking rate for particle
production when the calculation is interpreted quantum mechanically.

Unruh's model can be interpreted as a field theory in a two dimensional
black hole spacetime, without any reference to fluid flow.\footnote{The
corresponding spacetime is geodesically incomplete in somewhat the
same way the Eddington-Finkelstein coordinates are incomplete.
(For the details in a slightly different setting see 
Ref. \cite{JVfilm}.)
This is curious, since from the Newtonian point of view the fluid 
it must of course be 
{\it physically} complete. The Lorentzian incompleteness is physically
irrelevant since, as a wave propagates towards the edge of the spacetime,
it blueshifts beyond the linear, Lorentz-invariant 
part of the dispersion relation. Thus,
for example, a wave can not fall off the spacetime running backwards
in time along the horizon since
it will first blueshift and, in the subluminal case, undergo 
a reversal of group velocity which takes it back away from the horizon.
In the superluminal case it will cross the horizon.} 
As such, it becomes a model of how quantum fields may behave
in a black hole background if there is modified 
dispersion at high wavevectors. The modified dispersion relation
is not Lorentz invariant, so one must specify the local frame in
which the dispersion is specified. The spacetime analog of the 
co-moving fluid frame is the free-fall frame of the black hole,
and this frame has been adopted in most of the work to date.
As discussed in Section \ref{lorentz}, the results are independent
of the choice of preferred frame as long as it is not too 
accelerated or boosted relative to the black hole. 

Subsequent to 
Unruh's calculation much work has been done to understand, 
confirm, and extend the result. Here I will briefly review this
work. First, the result was explained by Brout et. al. \cite{Brout} 
using a WKB analysis, in which they computed the 
Bogoliubov coefficients at leading order. They also showed how
the trajectories of the wavepacket and its negative energy ``partner"
can easily be found from the geometric optics limit of the 
modified wave equation. 

The result was confirmed to very high precision by Corley
and Jacobson\cite{CJhfd}, by exploiting the stationarity to
reduce the problem to one of (numerically)
solving the ODE's for modes of fixed frequency
satisfying the appropriate (damped) boundary condition 
inside the horizon. In order to keep the order of the ODE's
from going higher than four derivatives, the dispersion
relation 
\begin{equation}
\omega^2=k^2-k^4/k_0^2
\label{quartic}
\end{equation}
was adopted, the idea
being that the $k^4$ term is just the lowest order term in the
derivative expansion of a generic (subluminal) dispersion 
relation.
This work also revealed a new mechanism
of particle creation, in addition to the Hawking radiation, in
which particles are created by static curvature which
however is not static in the frame in which the 
dispersion relation is specified. Corley\cite{Cn} further studied this
new type of particle creation in settings where there is no black
hole, and explored the modification of the Hawking spectrum in
the limit where the Hawking temperature approaches or exceeds the 
scale $k_0$ at which the dispersion relation is modified. 

Corley also developed\cite{Ca} the technology for analytical calculations
of the Hawking effect in dispersive field theory models, using the 
method of matched asymptotic expansions. In this method, one
solves the ODE for the mode functions 
by the method of Laplace transform in a neighborhood of the
horizon, and matches this solution to the WKB modes away from the 
horizon. This technology has previously
been well developed in the context of plasma 
wave theory\cite{Swanson}, where the
analogs of the wave phenomena we are discussing go by the name of 
``mode conversion". Using this method, Corley reproduced the leading
order Hawking result, and treated also the case
of superluminal propagation at high wavevectors, which is relevant 
to some condensed matter models (see below). He showed that
in the presence of superluminal high frequency dispersion 
the Hawking effect would also be recovered at
leading order, provided the modes approaching the horizon 
from the inside are in their ground state. 

Corley's methods have recently been extended 
by both Himemoto and Tanaka\cite{HT} and by 
Saida and Sakagami\cite{SS},
to find the leading order deviations from the thermal Hawking spectrum. 
In Ref. \cite{SS} a precise form for the leading deviation
is found, for frequencies in the range
$\kappa<\omega<\kappa(k_0/\kappa)^{2/5}$, where $\kappa$
is the surface gravity. This deviation is of order 
$\omega^3/\kappa k_0^2$, in agreement with the results of  
in Ref. \cite{HT} which show that around the 
peak of the Hawking spectrum, 
the corrections are generically of order
$\kappa^2/k_0^2$.
Both of these papers confirm this prediction with numerical 
calculations (although the sign of the corrections 
found numerically in Ref.\cite{SS} does not match the 
analytically computed
correction). 
This result disagrees with the 
earlier numerical results of Ref.\cite{CJhfd}, which showed an even smaller
deviation from the thermal result.
As shown in both these papers, however, 
this discrepancy is explained by the fact that the particular 
black hole metric adopted in Ref. \cite{CJhfd} was not generic
in its form near the horizon, and produced abnormally small deviations.

The analysis of \cite{Ca} was applied by Corley and 
Jacobson\cite{CJsuper} to the case, motivated by 
condensed matter models, 
where a field with superluminal high frequency dispersion propagates
in a black hole background with both an inner horizon and 
an outer horizon.
In this case it was found that the negative energy partners of 
Hawking quanta bounce off the inner horizon, return to the outer
horizon, and stimulate more Hawking radiation if the field is 
bosonic or suppress it if the field is
fermionic. This process leads to exponential growth or damping 
of the radiated flux and correlations
among the quanta emitted at different times, 
unlike in the usual Hawking effect.

An intriguing question that remains open in these dispersive models
is the behavior of Hawking radiation in the ultra low frequency
limit. There is some reason to suspect a strong
deviation from the thermal Hawking spectrum in this limit.
The WKB wavevector $k_{\rm tp}$ at the
turning point is of order\cite{CJlattice}
$k_{\rm tp} \sim \omega^{1/3}k_0^{2/3}$,
which is smaller than the surface gravity $\kappa$ when
$\omega<\kappa^3/k_0^2$, whereas the usual approximation
underlying the derivation of the Hawking spectrum 
assumes on the contrary that the wavevector near the
horizon is much larger than $\kappa$. 
On the other hand, perhaps conditions at the turning point 
are irrelevant. In fact, at least for frequencies of order 
$\kappa$ or greater, the Hawking spectrum 
seems to be determined by the behavior of the waves as they
tunnel across the horizon, as evidenced by 
the fact that the thermal spectrum at the Hawking temperature 
is recovered to high precision 
even when the turning point recedes significantly 
from the horizon.\cite{CJhfd,JMlattice}.  
Whether this `horizon dominance' persists in the 
ultra low frequency regime is not obvious.
It seems worthwhile to settle this question, as it would 
be fascinating and important if a modification of the theory at the 
high wavevector scale $k_0$ were to have an effect on the 
Hawking spectrum at ultra low frequencies.

\subsection{Revenge of the trans-Planckian modes}
As explained above in Section \ref{sonic}, 
the mechanism of producing the outgoing modes is a reversal of
group velocity of high wavevector (of order $k_0$) 
modes near the horizon, brought about 
by the nonlinearity of the dispersion relation $\omega(k)$. This 
avoids the need to draw upon a reservoir of trans-Planckian 
modes at the horizon. However, it does not fully solve the problem.
If we ask where these ingoing, short wavelength modes come from,
there is no satisfactory answer. In Unruh's model, with the 
$\tanh$ dispersion relation (\ref{tanh}), there is no cutoff at
high wavevectors, and the group velocity in the co-moving frame
drops to zero as $k\rightarrow\infty$. This means that as 
the ingoing mode is followed further out backwards in time,
it continues to blueshift without bound as the velocity 
of the co-moving frame (relative to the static frame) drops to zero.
Thus the need for an infinite density of states has not 
been eliminated, it has only been pushed away from the horizon out
to infinity. 

In the dispersive models with the quartic dispersion relation
(\ref{quartic}), the problem is different. As the ingoing
mode is followed backward in time, the wavevector
runs out to the end of the spectrum at $|k|=k_0$. 
(This happens to the positive frequency part only asymptotically
where the co-moving velocity drops to zero, however the
negative frequency part of the mode encounters the end of the
spectrum at non-zero co-moving velocity.\cite{CJhfd}) At that 
point the model is ill-behaved, since for $|k|>k_0$ the frequency
is imaginary. Of course the quartic dispersion relation was 
never intended as a model for a fundamental theory, but only 
as a first order correction, so it is not disturbing that it does not
make sense when pushed to sufficiently high wavevectors. 

Another way to see that it is impossible to sensibly produce the
outgoing modes in these models is to note that the Killing frequency
is conserved since the background is static. Thus the ingoing
mode must have the same frequency as the outgoing mode. However,
this implies that there is no Hawking radiation, since in the absence
of mixing between positive and negative frequencies the in-vacuum evolves
to the out-vacuum. 

In studying these models, this problem was avoided simply by 
imposing the ingoing vacuum boundary condition at non-zero 
co-moving velocity, never taking the asymptotic velocity to zero. 
In principle, however, the problem means that these models have 
not succeeded in providing a fully viable mechanism for producing 
the outgoing black hole modes. 

\section{Lattice models}
\label{lattice}

A lattice provides a simple model for imposing a physically 
sensible short distance cutoff, one which is more like the
atomic fluid of Unruh's sonic model than are the continuum-based
dispersive models. In a lattice model, space is discretized,
and a linear field takes values only at the lattice points,
each of which has a continuous worldline in spacetime. 
The discrete theory is self-contained
and unitary, so there can be no physically pathological 
behavior at high lattice wavevectors. 

The lattice succeeds in producing the outgoing modes
since, on a lattice of spacing $\delta$, 
the dispersion relation is
sinusoidal, wavevectors are identified modulo 
$2\pi/\delta$, and 
continuous passage from 
left-moving to right-moving wavepacket is possible.
Let me now explain this. The details are described in 
Refs. \cite{CJlattice,JMlattice}.

It might seem at first that
the most natural choice would be to preserve the time translation
symmetry of the spacetime with a static lattice whose points
follow accelerated worldlines. On a static lattice, however, the
Killing frequency is conserved, so outgoing modes arise from
ingoing modes with the same frequency, and there is no Hawking
radiation. The in-vacuum therefore evolves to a singular state at the
horizon. (The equilibrium (Hartle-Hawking) 
state at the Hawking temperature
would however presumably be regular at the horizon, hence could be
modeled on a static lattice. This thermal equilibrium state
could also be modeled on a static lattice in a Euclidean black hole 
spacetime\cite{lennyeuclid}.)
Moreover, inside the horizon the Killing field is
spacelike, so the worldlines of static lattice points would
be spacelike, which would make the lattice theory sick
if the inside were not omitted.
Finally, a static lattice is unnatural from the viewpoint
of the fluid model, wherein the atoms flow across the horizon.

If the lattice points are instead falling, 
it is still possible to preserve 
a discrete remnant of time translation symmetry.
A single falling worldline can be 
repeatedly translated in Killing time by a discrete amount,
building up a lattice. If the lattice points are asymptotically
at rest at infinity however, their spacing 
will go to zero at infinity, so there is no fixed
short distance cutoff.\footnote{Hawking radiation on such a lattice 
was investigated analytically in \cite{CJlattice}, by
exploiting the discrete symmetry to reduce the 2d lattice
wave equation to a 1d difference equation. This made it
possible, with the help of the methods of \cite{Ca},
to analytically establish the leading order Hawking effect.}

We thus insist that the lattice spacing asymptotically
approaches a fixed constant {\it and} that the lattice points 
are at rest at infinity and fall freely into the black hole. 
In this case, 
the lattice cannot have even a discrete time
translation symmetry, since there is a gradual spreading
of the lattice points as they fall  
toward the horizon. The time scale
of this spreading is of order $1/\k$ where $\k$ is the surface
gravity. This time dependence of the
lattice is invisible to long wavelength modes, which sense
only the stationary background metric of the black hole, but it is
quite apparent to modes with wavelengths of order the lattice spacing.

On such a lattice the long wavelength outgoing modes 
come from short wavelength ingoing modes via a process
closely analogous to the Bloch oscillation of an accelerated
electron in a crystal. Bloch oscillations occur
when the acceleration acts long enough for
the momentum to grow to the scale of the lattice wavevector.
Due to the sinusoidal nature of the dispersion relation for the
electron in the lattice,
the group velocity drops, and changes sign
when the momentum reaches $\pi/\delta$, where 
$\delta$ is the lattice spacing.
An accelerated electron would therefore oscillate back and forth
(were the motion not dissipated by coupling to other
lattice degrees of freedom such as phonons).

On a falling lattice in a black hole background, something
similar happens. In a freely falling, 
Gaussian normal coordinate system, the 1+1 dimensional
black hole metric takes the form 
\begin{equation}
ds^2 = dt^2 - a(z,t) dz^2,
\end{equation}
where the ``scale factor" $a(z,t)$ goes to unity at $t=0$
and as $z\rightarrow \infty$.
If the $z$-coordinate is discretized as $z_n = n\delta$,
the dispersion relation for a massless scalar field 
mode $\exp(-i\omega t + ikz_n)$ on the 
lattice takes the form 
\begin{equation}
\omega = \pm \frac{2}{a(z,t)\, \delta}\sin(k\delta/2).
\end{equation}
This is the standard lattice dispersion relation,
with the additional factor $1/a(z,t)$ coming from
the black hole geometry. 

Following an outgoing wavepacket backwards towards
the horizon, the gravitational field blueshifts the 
wavevector as in the continuum. 
In a WKB worldline approximation, we can 
describe what happens backwards in time as follows.
The group velocity drops below the 
velocity of lattice points near the horizon, so the 
wavepacket turns around with respect to the static
coordinate and heads away from the horizon. At this stage,
however, the wavepacket is still outgoing with respect 
to the lattice, and its WKB wavelength can still be rather
long compared with the lattice spacing. The analog
of Bloch oscillation has not yet occurred. In this part of
the process the Killing frequency is conserved
since the wavepacket cannot sense the time dependence
of the lattice.  As the wavepacket propagates 
backwards in time away from the horizon
however, the blueshifting continues, eventually pushing the 
wavevector over the hump in the dispersion relation, so
the group velocity changes sign with respect to the 
lattice as well. At this stage the wavelength is
comparable to the lattice spacing, so the Killing frequency
is no longer conserved. Finally, at early times far 
from the horizon, the ingoing wavepacket which
produces the final outgoing wavepacket still has 
wavevector and frequency of order the
lattice spacing.

This ingoing wavepacket not only has a different  
frequency than the outgoing wavepacket, it has negative frequency
components as well. The presence of the negative frequency 
part is the signal of the Hawking effect. The appearance
of the negative frequency part in this context
can be understood 
as a consequence of conservation of degrees of freedom
in the following hand-waving manner.
As described so far, 
the entire outgoing branch of the dispersion
curve seems to be produced
from only a part of the ingoing branch, 
\begin{equation}
(-\pi/\delta, -k_c)
\quad\rightarrow\quad (0,\pi/\delta)
\end{equation}
where $k_c$ is some critical wavevector separating
the ordinary ingoing modes that cross the horizon from the
exotic ones that undergo the Bloch oscillation.
This cannot be the whole story, 
however, since there appear to be 
more outgoing modes than ingoing modes which give rise to them.
The resolution is that we left out the negative frequency
part of the ingoing mode. 

This WKB picture on the lattice, largely developed in \cite{CJlattice},
was checked in \cite{JMlattice}
by carrying out the exact calculation numerically 
using the lattice wave equation. The behavior of 
wavepackets described here was confirmed,  
and the Hawking radiation was recovered 
to within half a percent for a lattice spacing
$\d=0.002/\k$ (which corresponds to a proper spacing 
$a(z,t)\delta\sim 0.08/\k$ 
where the wavepacket turns around at the horizon). 
The deviations
from the Hawking effect that arise as $\kappa\d$ is increased
were also studied, and an interesting picture of how the wavepackets
turn around at the horizon was revealed.

\subsection{Towards a non-expanding lattice model:
back-reaction and dissipation?}

The falling lattice model has provided an intriguing
mechanism---the Bloch oscillation---for getting an outgoing
mode from an ingoing mode in a stationary background, but there
is a serious flaw in the picture: the lattice is constantly
expanding. In the fluid analogy, 
by contrast, the lattice of atoms maintains
a uniform average density. In a fundamental theory we might also
expect the scale of graininess of spacetime to remain fixed 
at, say, the Planck scale or the string scale (since presumably
the graininess would {\it define} this scale) rather than 
expand. Can the falling lattice model be improved to share this
feature? 

An incompressible fluid maintains uniform density in an 
inhomogeneous flow by compressing in some directions and 
expanding in others, a  process requiring at least two dimensions.
At the atomic level such a volume-preserving
flow involves erratic motions of individual atoms. One possible
improvement of the lattice model is to make a two-dimensional lattice
that mimics this sort of volume preserving flow. It is not 
clear whether the motions of the lattice points can be slow
enough to be adiabatic on the time scale of the high frequency
lattice modes. If they cannot, then the time dependence of 
this erratic lattice background will excite the quantum vacuum.

In a fluid, however, 
the lattice is a part of the system, not just a fixed
background. The physical ground state requirement is that 
the time-dependence of the flow be adiabatic 
for the fully coupled system.  A similar comment applies in quantum
gravity: surely, if in-out mode conversion is at play, the incoming 
high frequency modes are strongly coupled to the quantum gravitational
vacuum. Ideally, therefore, we should try to find a model in which
the background is not decoupled from the perturbations.
The analogy with the dissipation of Bloch oscillations due to
coupling to the lattice phonon modes strongly suggests that,
in such a nonlinear model, the outgoing modes will arise from
the time reverse of a dissipative process. The same 
idea was suggested from a very different point of view 
by Brout et al\cite{Brout2}, who examined the Hawking process 
in a dissipative, dispersive effective field theory derived
from a modification of quantum commutators motivated by 
string theory. It is also consistent with the observation of 
Visser\cite{Visser} that the addition of a viscosity term to
the Euler equation for a fluid results in a dispersion relation
for perturbations similar to (\ref{quartic}) but with an imaginary
part: $\omega = (k^2 - k^4/k_0^2)^{1/2} - ik^2/k_0$.
(In this case, when the dispersion
comes entirely from viscosity, the 
imaginary correction is larger than the real one by a factor
$k_0/k$.)

A first step in this direction might be to study a
one dimensional quantum ``chain"
model in which the lattice points are non-relativistic
point masses, coupled to each other by nearest neighbor interactions, 
and ``falling" or propagating in a background potential (with
or without periodic boundary conditions) with asymptotically
non-zero velocity.
The perturbations of
such a chain are the phonon field, and the back reaction to the 
Hawking radiation is included
(although the background potential is fixed). 
In a model like this one could presumably follow in detail the 
nonlinear origin of the outgoing modes and the transfer of 
energy from the mean flow to the thermal radiation.

The simplest such chain model would be one with harmonic nearest
neighbor interactions. However, as pointed out by 
Unruh\cite{billharmonic}, a harmonic chain cannot possess
a horizon, since the sound velocity $v_s$ is proportional to the 
inter-atomic spacing, as is the flow velocity $v(x)$ in a stationary
flow (because the mass current is uniform). 
When the chain is stretched in a region of higher flow velocity, 
the ratio $v_s(x)/v(x)$ therefore remains constant, so no
horizon forms. This is unfortunate, since the harmonic chain
could have been solved exactly, at least asymptotically.
Perhaps the idea would work with another exactly solvable
model, such as the Calogero-Sutherland model\cite{CSmodel},
which consists of nonrelativistic point masses coupled by 
$1/x^2$ interactions. Of course this
model is exactly solvable only in isolation, not in a background
potential such as needed to accelerate the chain near the
horizon. Nevertheless, the asymptotic exact solution may be useful.
Alternatively, perhaps no exact solution is needed at all.

\section{Condensed matter analogs}

The title of Unruh's original paper\cite{Unruh81} on the sonic
analogy was ``Experimental black-hole evaporation?" He was not 
only proposing a fluid model in which the effects of a 
short distance cutoff and quantum back-reaction to the 
Hawking radiation
should be more understandable than in quantum gravity,
he was also suggesting that these processes might be 
experimentally observable in the analog fluid system.

The Hawking effect is a quantum process involving an
instability of the ground state due to the presence of the 
ergoregion behind the horizon. If a fluid model is to 
produce identifiable Hawking radiation, therefore, the flow should 
be in its quantum ground state (or at least be as cold as the Hawking
temperature), rather than in an incoherent 
thermal state. For this reason we should contemplate setting
up a horizon in a superfluid at zero temperature. 
The case of superfluid ${}^4$He
was initially examined in \cite{ultrashort}, and further
discussed in \cite{TJorigin}.
It was concluded that a sonic horizon cannot be 
established in superflow, because the flow is unstable
to roton creation at the Landau velocity which is some
four times smaller than the sound velocity. 

There may be other condensed matter systems, however, where
a Hawking effect analog can be observed. One system that
has been 
studied\cite{JVhorizons,KVdecay,JVfilm,Vtorus} 
is the (anisotropic) A-phase of superfluid 
${}^3$He\cite{VollWolf,Vexotic},
which has a rich spectrum of massless 
quasiparticle excitations. In particular, there are fermionic
quasiparticles---the ``dressed" helium atoms---which have
gapless excitations near the gap nodes at $\vec{p}=\pm p_F\hat{l}$
on the anisotropic
Fermi surface, and therefore can play the role of a massless
relativistic field in a black hole analog. The unit vector
$\hat{l}$ is the direction of orbital angular momentum
of the $p$-wave Cooper pairs and $p_F$ is the Fermi momentum. 

The velocity of fermion quasiparticles parallel
to $\hat{l}$ in ${}^3$He-A 
is the Fermi velocity $v_F\sim 55$ m/s, while their
velocity perpendicular to $\hat{l}$ is only 
$c_\perp =\Delta/p_F\sim 3$ cm/s, where $\Delta\sim T_c\sim 1$ mK 
is the energy gap. It should be possible to set
up an inhomogeneous superflow exceeding the slow speed $c_\perp$
in a direction normal to $\hat{l}$, thus creating a horizon
for the fermion quasiparticles. There is a catch, 
however, since the superflow is unstable when 
the speed relative to a container exceeds $c_\perp$.\cite{KVdecay}.
A possible way around this was suggested by Volovik\cite{Vtorus},
who considered a thin film of ${}^3$He-A flowing on a substrate
of superfluid ${}^4$He, which insulates the ${}^3$He from 
contact with the container. He imagined a radial flow
on a torus, such that the flow velocity near the inner 
radius exceeds $c_\perp$, producing a horizon. 
Theoretically this looks promising,
however the Hawking temperature for a torus of size $R$ is
$T= (\hbar/2\pi)(dv/dr)\sim \hbar c_\perp/R\sim (\lambda_F/R)$ mK,
where $\lambda_F$ is the Fermi wavelength, which is of 
the order of  
Angstroms. Thus, even for a micron sized torus, the Hawking temperature
would be only $\sim 10^{-7}$ K.

An alternative is to keep the superfluid at rest with respect to 
the container, but arrange for a texture in the order parameter
to propagate in such a way as to create a horizon. For example,
in \cite{JVhorizons} a moving ``splay soliton" is considered.
This is a planar texture in which the $\hat{l}$ vector 
rotates from $+\hat{x}$ to $-\hat{x}$  along
the $x$-direction perpendicular to the soliton plane.
A quasiparticle moving in the $x$-direction thus goes
at speed $v_F$ far from the soliton and at speed $c_\perp$
in the core of the soliton. If the soliton is moving at
a speed greater than $c_\perp$, the quasiparticles will not
be able to keep up with it, so an effective horizon will
appear. This example turns out to be rather interesting and 
complicated 
in the effective relativistic description. 
The horizon has a translational velocity,
making it like that of a rotating black hole rather than
a static black hole. In addition, there is a strong
pseudo-electromagnetic field outside the ``black hole", which 
would produce quasiparticles by pseudo-Schwinger pair 
production.\cite{Vexotic} (This latter process may be the same as 
what produces the so-called ``orbital viscosity"\cite{VollWolf}
of a time-dependent texture.) The Hawking temperature also tends to
be very low, and it seems likely that the Hawking effect would be
masked by the pseudo-Schwinger pair 
production, though this has not been definitively analyzed.

In \cite{JVfilm} a simpler system was studied, that of 
a thin film of ${}^3$He-A, perhaps on a ${}^4$He substrate,
with a domain wall in which the condensate is in a different
superfluid phase
and across which the direction of $\hat{l}$,
which is perpendicular to the film, flips sign. Inside the
wall the group velocity of the quasiparticles goes to zero,
so if the wall itself is propagating, a horizon
will appear. The effective spacetime geometry of this system
was studied in \cite{JVfilm}, and it is a potentially interesting
black hole candidate. However, the presence of the moving domain wall
raises questions about the evolution of the quasiparticle vacuum
that have not been addressed. Moreover, in this as in any
model in which the ergoregion has a finite extent, bounded
on the inside by a white hole or inner horizon, it would
be necessary to understand the time scale on which the 
filling of the negative
energy states in the ergoregion would turn off the Hawking
process. In a black hole, by contrast, the negative
energy states just fall into the singularity, never to 
be heard from again.

Turning away from ${}^3$He-A, 
some other systems have been considered
as candidates for black hole analogs in condensed matter.
A non-axisymmetric vortex in ${}^3$He-B has gapless
excitations in the states bound to the vortex core,
and if the core is rotating this can lead to a black hole
analog for these modes, as discussed by 
Kopnin and Volovik\cite{KVvortex}.
Reznik\cite{Reznik} discussed a model involving a 
dielectric medium 
with spatially varying index of refraction.
Hochberg and P\'erez-Mercader\cite{Hoch} developed a liquid 
model for black hole thermodynamics (but not for the Hawking effect).
Most recently, the possibility of realizing a 
sonic black hole analog in a dilute 
Bose-Einstein condensate has been discussed.\cite{Anglin}
The condensate could possibly be made to flow in a circulating  
manner, with a constriction leading to a black hole/white hole 
horizon pair with an ergoregion in between. The dispersion relation
for sound in this system 
is ``superluminal" at large wavevectors, so the system
falls into the class of models shown to be unstable to 
a runaway process of stimulated emission of Hawking 
radiation.\cite{CJsuper} The analysis of Ref. \cite{CJsuper} was based 
on a study of WKB wavepackets, which may not be a valid approximation
anywhere in the regime of interest, however a numerical investigation
of the linearized modes of the Gross-Pitaevskii equation appears
to bear out the same conclusion.\cite{Anglin}

We must now leave the topic of condensed matter analogs 
with many open questions. Hopefully one day some
systems in which a Hawking effect can be observed will
be identified.
 
\section{Lorentz non-invariance}
\label{lorentz}
A short distance cutoff or modified dispersion relation
requires a breaking of local Lorentz invariance
(or perhaps a breakdown of locality itself),
since the distinction between long and short wavelengths
depends on the frame of reference.
In the models discussed so far, this frame was taken
to be the one defined by geodesics that are asymptotically at 
rest at infinity and fall across the horizon---the ``free-fall frame".
In terms of the unit velocity 2-vector $u$ and the Killing vector
$\xi$, this frame is specified by the unit energy condition 
$u\cdot\xi=1$. 
The rate of change of $\xi^2$ along the free-fall
worldlines characterizes the time-dependence of the 
black-hole background seen from the point of view of the
free-fall frame. At the horizon this rate takes the value $-2\kappa$,  
so the time scale associated with the free-fall motion
is of the same order as that defined by the surface gravity $\kappa$.
It is plausible, however,  that as long as the 
choice of preferred frame does not introduce a time scale 
comparable to $1/k_0$, 
the results should not depend on this choice at leading order.

The dependence of the results on the choice of preferred frame
has been studied by Himemoto and Tanaka\cite{HT}.
They consider accelerated frames (which are in fact the free-fall 
frames of 
metrics related to the original metric by a static conformal factor).
They find analytically that, as long as the acceleration
of the frame is 
not too drastic, the leading deviation from the thermal spectrum
occurs at order $1/k_0^2$. They also investigated numerically the case 
where the acceleration becomes large, so the frame is prevented
more and more from falling freely, and found that the created
particle flux drops significantly, by something of order unity.
They conjecture that in the case of a static frame, which is infinitely
accelerated at the horizon, there may be no Hawking radiation at all.
A similar observation was made in the context of a lattice model in 
Ref. \cite{JMlattice}, for the case where the lattice
points follow static worldlines. It was
pointed out there that if the lattice points are static, then Killing 
frequency must be conserved on the lattice, so there is no possibility
of a positive frequency wavepacket developing negative frequency
components, which rules out any Hawking effect. This means that,
on such a lattice, the in-vacuum must evolve to the Boulware vacuum
at the horizon. 
Thus, while it is not critical to the Hawking effect 
that the free-fall frame be adopted, the preferred frame
should not be too drastically accelerated. 

If we are to entertain the possibility of a true breaking
of Lorentz invariance in Nature, it seems that the asymptotic
rest frame of a black hole would be the preferred one only to the 
extent that the black hole is at rest with respect to 
the cosmic preferred frame, whatever that is. As just discussed,
however, it should not be important that the black hole be precisely 
at rest, but just that the relative boost factor $\gamma$ between 
the black hole and the cosmic preferred frame be much smaller than
$k_0/\kappa$. Only for a near-Planck mass primordial black hole in the 
very early universe is it conceivable that this restriction would 
be violated.

In a Robertson-Walker cosmology  
the cosmic rest frame would plausibly be 
the frame of the isotropic observers. 
In a cosmology with less symmetry, there 
is presumably no simple way to characterize the cutoff precisely.
However, it is not implausible that in a rough approximation
a preferred frame is given by the level sets of the 
cosmological time function\cite{CtimeA,CtimeW}, 
i.e. the length of the longest
timelike curve back to the initial singularity (or 
back to some initial
slice in the quantum gravity era). Although this time function
is not smooth, its first and second derivatives exist almost 
everywhere\cite{CtimeA}, so in particular 
its gradient defines a local frame almost everywhere.
It may be possible to construct a ``phenomenological" 
theory using this
cosmic rest frame to specify some new physics 
at short distances\cite{JMctime}. 
In such a framework,
local Lorentz invariance is broken while preserving general
covariance, since the preferred local frame is not ``additional furniture"
but rather is determined (non-locally) by the metric. The gravitational 
coupling of the new physics is therefore determined by its metric-dependence.

\section*{Acknowledgements}
I am grateful to my collaborators
Steve Corley, Dave Mattingly, and Grisha Volovik,
as well as to Bill Unruh,
for countless instructive and informative
discussions on the matters described here. 
This work was supported in part by the 
National Science Foundation
under grants No. PHY98-00967 
at the University of Maryland and PHY94-07194 
at the Institute for 
Theoretical Physics.

%\appendix
%\section{First Appendix} %Empty argument \section{} yields `Appendix'. 

%\section{Second Appendix}


\begin{thebibliography}{99}
\bibitem{rho} This point was not discussed in Ref. \cite{Weiss}.
\bibitem{lennyeuclid}The possibility of computing
thermal field effects on a euclidean black hole lattice 
was suggested to me by L. Susskind.
\end{thebibliography}
\end{document}